\documentclass{amsart}
\newcommand{\R}{\mathbb{R}}
\def\x{{\vec{x}}}
\def\E{{\mathcal E}}
\newcommand{\al}{\alpha}
\newcommand{\suli}{\sum\limits}
\newcommand{\rmax}{\rho_{\al,{\rm max}}}
\newcommand{\rmin}{\rho_{\al,{\rm min}}}

\newtheorem{theorem}{Theorem}[section]
\newtheorem{lemma}[theorem]{Lemma}

\begin{document}
%------------------------------------------------------------
% Titlepage
%------------------------------------------------------------

\title[Bosons in a Trap]{Bosons in a Trap: Asymptotic Exactness of the 
Gross-Pitaevskii Ground State Energy Formula}
% The title of your article 

%\titlerunning{Bosons in a Trap}
% allows abbreviation of title, if the full title is too long
% to fit in the running head

\author{Robert Seiringer}
\address{Institut f\"ur Theoretische Physik,
Universit\"at Wien, Boltzmanngasse 5, A-1090 Vienna,
Austria}
\email{rseiring@ap.univie.ac.at}

%------------------------------------------------------------
% Subject classifications
%------------------------------------------------------------


%------------------------------------------------------------
% End of document
%------------------------------------------------------------

%\authorrunning{Robert Seiringer}
% if there are more than two authors,
% please abbreviate author list for running head

%------------------------------------------------------------
% Abstract. 
%------------------------------------------------------------

\begin{abstract}
Recent experimental breakthroughs in the treatment of dilute Bose
gases have renewed interest in their quantum mechanical
description, respectively in approximations to it. The ground
state properties of dilute Bose gases confined in external
potentials and interacting via repulsive short range forces are
usually described by means of the Gross-Pitaevskii energy
functional. In joint work with Elliott H. Lieb and Jakob Yngvason 
its status as an approximation for the quantum
mechanical many-body ground state problem has recently been
rigorously clarified. We present a summary of this work, for both
the two- and three-dimensional case.
\end{abstract}

%------------------------------------------------------------
\maketitle
%------------------------------------------------------------

\section{Introduction}

The Gross-Pitaevskii (GP) functional was introduced in the early
sixties as a phenomenological description
of the order parameter in superfluid ${\rm He}_{4}$ \cite{G1961,P1961,G1963}. 
It has come
into prominence again because of recent experiments on
Bose-Einstein condensation of dilute gases in magnetic traps. The
paper \cite{DGPS} brings an up to date review of these
developments.

The present contribution is based on the joint work \cite{LSY1999,LSY2000} with 
Elliott H. Lieb and Jakob Yngvason (see also \cite{LSY1999a}). 
The starting point of our investigation is the Hamiltonian  for
$N$ identical bosons moving in $\R^D$, $D=2$ or $3$, that interact
with each other via a radially symmetric pair-potential $v(|\x_i -
\x_j|)$ and are confined by an external potential $V(\x)$:
\begin{equation}
H = \sum_{i=1}^{N} \{- \Delta_i + V(\x_{i})\}+
\sum_{1 \leq i < j \leq N} v(|\x_i - \x_j|).
\end{equation}
The Hamiltonian acts on {\it symmetric} wave functions in
$\otimes^N L^2(\R^{D},d\x)$.  The pair interaction $v$ is assumed
to be {\it nonnegative} and of short range, more precisely, we
demand it to have a finite scattering length. (For a definition of
the scattering length in arbitrary dimension see \cite{LY2000}.)
The potential $V$ that represents the trap is locally bounded and
$V(\x)\to\infty$ as $|\x|\to\infty$. By shifting the energy scale
we can assume that $\min_{\x}V(\x)=0$.

Units are chosen so that $\hbar=2m=1$, where $m$ is the particle
mass. A natural energy unit is given by the ground state energy
$\hbar\omega$ of the one particle Hamiltonian
$-(\hbar^2/2m)\Delta+V$. The corresponding length unit,
$\sqrt{\hbar/(m\omega)}$, measures the effective extension of the
trap.

We are interested in the ground state energy $E^{\rm QM}=\inf{\rm
\,spec\,}H$. Besides $N$ it depends on the potentials $V$ and $v$,
but with $V$ fixed and
\begin{equation}\label{scalv}
v(r)=(a_1/a)^2v_1(a_1r/a),
\end{equation}
where $v_{1}$ has scattering length $a_{1}$ and is regarded as
{\it fixed}, $E^{\rm QM}$ is a function of $N$
and $a$ only. The corresponding ground state density is given by
\begin{equation}
\rho^{\rm
QM}(\x)=N\int|\Psi_{0}(\x,\x_{2},\dots,\x_{N})|^2d\x_{2}\dots
d\x_{N},
\end{equation}
where $\Psi_0$ is a ground state wave function of $H$.

Note that $v$ given in ({\ref{scalv}) has scattering length $a$.
Here $a$ is dimensionless and really stands for
$a\sqrt{m\omega/\hbar}$. Hence a scaling of $v$ like
(\ref{scalv}) is equivalent to scaling the external potential $V$
at fixed $v$. In particlar the limit $a\to 0$ with fixed $V$ is equivalent
to the limit $\omega\to 0$ with fixed $v$, if one introduces the 
scaling $V(\x)=\omega V_1(\omega 
^{1/2} \x)$ for some fixed $V_1$.

Recent experiments on Bose-Einstein condensation are usually
interpreted in terms of a function $\Phi^{\rm GP}(\x)$ of
$\x\in{\R}^D$, which minimizes the {\it Gross-Pitaevskii energy
functional}
\begin{equation}\label{gpf}
\E^{\rm GP}[\Phi]=\int_{\R^D}\left(|\nabla\Phi|^2+V|\Phi|^2+4\pi
g|\Phi|^4\right)d\x
\end{equation}
under the subsidiary condition $\int|\Phi|^2=N$. The corresponding
energy is
\begin{equation}
E^{\rm GP}(N,g)=\inf_{\int|\Phi|^2=N}\E^{\rm GP}[\Phi]=\E^{\rm
GP}[\Phi^{\rm GP}].
\end{equation}
The parameter $g$ is different in dimensions 2 and 3. However, for
any value of $g>0$ and $N>0$ it can be shown that a unique,
strictly positive $\Phi^{\rm GP}$ exists \cite{LSY1999}. It
depends on these parameters, of course, and when this is important
we denote it by $\Phi^{\rm GP}_{N,g}$.

The motivation of the term $4\pi g|\Phi|^4$ in the GP functional
comes from the ground state energy density, $\varepsilon_0(\rho)$, of a a
dilute, thermodynamically infinite, homogeneous Bose gas of
density $\rho$, interacting via a repulsive potential with
scattering length $a$. The formulas for this quantity are older
than the GP functional \cite{BO,Lee,Schick}, at least for $D=3$,
but they have only very recently been derived rigorously for
suitable interparticle potentials. See \cite{LY1998} and
\cite{LY2000}. They are given by
\begin{eqnarray}\nonumber
\varepsilon_0(\rho)&\approx& 4\pi a \rho^2\qquad {\rm for\ }D=3, \\
\varepsilon_0(\rho)&\approx& 4\pi \rho^2 |\ln(a^2\rho)|^{-1}\qquad {\rm for\
}D=2,
\end{eqnarray}
where $\approx$ means that the formulas are valid for {\it dilute}
gases, where $a^D\rho\ll 1$. Hence the natural choice of the
parameter $g$ is
\begin{eqnarray}
g &=& a \qquad {\rm for\ }D=3, \\ \label{g2} g &=&
|\ln(a^2\bar\rho)|^{-1}\qquad {\rm for\ }D=2,
\end{eqnarray}
where $\bar\rho$ is the {\it mean GP density}
\begin{equation}\label{rhobar}
\bar\rho=\frac 1N\int|\Phi^{\rm GP}(\x)|^4 d\x.
\end{equation}
Note that $\Phi^{\rm GP}$ depends on $g$, so (\ref{g2}) together
with (\ref{rhobar}) are non-linear equations for $g$.
Alternatively, one could define $g$ using the minimizer for $g=1$
in the definition of $\bar\rho$. Since $\bar\rho$ appears only
under a logarithm, this would not effect our leading order
calculations. For the same reason one could use the TF minimizer 
(see below) instead of
the GP minimizer to define $g$. Note also that unlike in the
three-dimensional case $g$ depends on $N$ in the two-dimensional case.

The idea is now that with this choice of $g$ one should, for
{\it dilute} gases, have that
\begin{equation}\label{approx}
E^{\rm GP} \approx E^{\rm QM}\quad{\rm and}\quad \rho^{\rm
QM}(\x)\approx \left|\Phi^{\rm GP}(\x)\right|^2\equiv \rho^{\rm
GP}(\x).
\end{equation}
This is made precise in the following theorems. Note that by
scaling
\begin{equation}
E^{\rm GP}(N,g)=NE^{\rm GP}(1,Ng)\quad\mbox{and}\quad \Phi^{\rm
GP}_{N,g}(\x)=N^{1/2}\Phi^{\rm GP}_{1,Ng}(\x). \label{scaling}
\end{equation}
Hence $Ng$ is the natural parameter in GP theory. With this in
mind we can state our first main result.

\begin{theorem}[The GP limit of the QM ground state energy and
density]\label{thm1}  ~ \\ If $N\to\infty$ with $Ng$ fixed, then
\begin{equation}\label{econv}
\lim_{N\to\infty}\frac{{E^{\rm QM}(N,a)}}{ {E^{\rm GP}(N,g)}}=1,
\end{equation}
and
\begin{equation}\label{dconv}
\lim_{N\to\infty}\frac{1}{ N}\rho^{\rm QM}(\x)= \left |{\Phi^{\rm
GP}_{1,Ng}}(\x)\right|^2
\end{equation}
in the weak $L_1$-sense.
\end{theorem}

Note that by hypothesis of the theorem above 
it really applies to dilute gases, since
for fixed $Ng$ (which we refer to as the GP case) the mean density
$\bar\rho$ is of order $N$ and
\begin{equation}
a^3\bar\rho\sim N^{-2}\quad\mbox{for}\quad D=3,\quad
a^2\bar\rho\sim \exp(-N)\quad\mbox{for}\quad D=2.
\end{equation}
Especially for $D=2$ this is an unsatisfactory restriction, since
$a$ has to decrease exponentially with $N$. For a slower decrease
$Ng$ tends to infinity with $N$, and the same holds for $D=3$ if $a$
does not decrease at least as $N^{-1}$. In this case, the gradient term
in the GP functional becomes negligible compared to the 
other terms and the so-called {\it
Thomas-Fermi (TF) functional}
\begin{equation}\label{gtf}
\E^{\rm TF}[\rho]=\int_{\R^D}\left(V\rho+4\pi g\rho^2\right)d\x
\end{equation}
arises. It is defined for nonnegative functions $\rho$ on $\R^D$.
Its ground state energy $E^{\rm TF}$ and density $\rho^{\rm TF}$ are
defined analogously to the GP case. Our second main result is that
minimization of (\ref{gtf}) reproduces correctly the ground state
energy and density of the many-body Hamiltonian in the limit when
$N\to\infty$, $a^D\bar \rho\to 0$, but $Ng\to \infty$ (which we
refer to as the TF case), provided the external potential is 
reasonably well behaved. 
We will assume that $V$ is asymptotically equal to some
function $W$ that is homogeneous of some order $s>0$ and locally
H\"older continuous (see \cite{LSY2000} for a precise definition).
This condition can be relaxed, but it seems adequate for most
practical applications and simplifies things considerably.

\begin{theorem}[The TF limit of the QM ground state energy and
density]\label{thm2} ~ \\ Assume that $V$ satisfies the conditions
stated above. If $\gamma\equiv Ng\to\infty$ as $N\to\infty$, but
still $a^D\bar\rho\to 0$, then
\begin{equation}\label{econftf}
\lim_{N\to\infty}\frac{E^{\rm QM}(N,a)} {E^{\rm TF}(N,g)}=1,
\end{equation}
and
\begin{equation}\label{dconvtf}
\lim_{N\to\infty}\frac{\gamma^{D/(s+D)}}{N}\rho^{\rm
QM}(\gamma^{1/(s+D)}\x)= \tilde\rho^{\rm TF}(\x)
\end{equation}
in the weak $L_1$-sense, where $\tilde\rho^{\rm TF}$ is the
minimizer of the TF functional under the condition $\int\rho=1$,
$g=1$, and with $V$ replaced by $W$.
\end{theorem}

\noindent{\it Remark.} The theorems are independent of the
interaction potential $v_1$ in (\ref{scalv}). This means that in
the limit we consider only the scattering length 
effects the ground state properties, and not the details of
the potential. Note also that the particular limit we consider is
{\it not} a mean field limit, since the interaction potential is
very hard in this limit; 
in fact the term $4\pi g|\Phi|^4$ is mostly kinetic
energy.
\medskip

In the following, we will present a brief sketch of the proof of
Theorems \ref{thm1} and \ref{thm2}. We will derive appropriate
upper and lower bounds on the ground state energy $E^{\rm QM}$.
The convergence of the densities follows from the convergence of
the energies in the usual way by variation with respect to the
external potential. We refer to \cite{LSY1999} and \cite{LSY2000}
for details.

\section{Upper bound to the QM energy}

To derive an upper bound on $E^{\rm QM}$ we use a generalization
of a trial wave function of Dyson \cite{dyson}, who used this
function to give an upper bound on the ground state energy of the
homogeneous hard core Bose gas. It is of the form
\begin{equation}\label{ansatz}
\Psi(\x_{1},\dots,\x_{N})
    =\prod_{i=1}^N\Phi^{\rm
GP}(\x_{i})F(\x_{1},\dots,\x_{N}),
\end{equation}
where $F$ is constructed in the following way:
\begin{equation}F(\x_1,\dots,\x_N)=\prod_{i=1}^N
f(t_i(\x_1,\dots,\x_i)),\end{equation} where $t_i =
\min\{|\x_i-\x_j|, 1\leq j\leq i-1\}$ is the distance of $\x_{i}$
to its {\it nearest neighbor} among the points
$\x_1,\dots,\x_{i-1}$, and $f$ is a  function of $r\geq 0$. We
choose it to be
\begin{equation}
f(r)=\left\{\begin{array}{cl} f_{0}(r)/f_0(b) \quad &\mbox{for}\quad r<b\\
1 &\mbox{for}\quad r\geq b,
\end{array}\right.
\end{equation}
where $f_0$ is the solution of the zero-energy scattering equation
(see \cite{LY2000}) and $b$ is some cut-off parameter of order
$b\sim \bar\rho^{-1/D}$. The function (\ref{ansatz}) is not
totally symmetric, but for an upper bound it is nevertheless an
acceptable test wave function since the bosonic ground state
energy is equal to the {\it absolute} ground state energy.

The result of a somewhat lengthy computation is the upper bound
\begin{equation}\label{ubd}
E^{\rm QM}(N,a)\leq E^{\rm GP}(N,g)\times \left\{\begin{array}{ll}
1+O(a\bar\rho^{1/3}) \quad &\mbox{for}\quad D=3 \\
1+O(|\ln(a^2\bar\rho)|^{-p}) \quad &\mbox{for}\quad D=2,
\end{array}\right.
\end{equation}
with the power $p$ equal to 1 in the GP case and $s/(s+2)$ in the
TF case (where $V$ is asymptotically homogeneous of order $s$).

\section{Lower bound to the QM energy}

To obtain a lower bound for the QM energy the strategy is to
divide space into boxes and use the estimate on the homogeneous
gas, given in \cite{LY1998} and \cite{LY2000}, in each box with
{\it Neumann} boundary conditions. One then minimizes over all
possible divisions of the particles among the different boxes.
This gives a lower bound to the energy because discontinuous wave
functions for the quadratic form defined by the Hamiltonian are
now allowed. We can neglect interactions among particles in
different boxes because $v\geq 0$. Finally, one lets the box size
tend to zero. However, it is not possible to simply approximate
$V$ by a constant potential in each box. To see this consider the
case of noninteracting particles, i.e., $v=0$ and hence $a=0$.
Here $E^{\rm QM}=N\hbar\omega$, but a `naive' box method gives
only 0 as lower bound, since it clearly pays to put all the
particles with a constant wave function in the box with the lowest
value of $V$.

For this reason we start by separating out the GP wave function in
each variable and write a general wave function $\Psi$ as
\begin{equation}
\Psi(\x_{1},\dots,\x_{N})=\prod_{i=1}^N\Phi^{\rm
GP}(\x_{i})F(\x_{1},\dots,\x_{N}).
\end{equation}
This defines $F$ for a given $\Psi$ because $\Phi^{\rm GP}$ is
everywhere strictly positive, being the ground state of the
operator $- \Delta + V+8\pi g|\Phi^{\rm GP}|^2$. We now compute
the expectation value of $H$ in the state $\Psi$. Using partial
integration and the variational equation for $\Phi^{\rm GP}$, we
see that for computing the ground state energy of $H$ we have to
minimize the normalized quadratic form
\begin{equation}\label{ener3}
Q(F)=\suli_{i=1}^{N} \frac{\int\prod_{k=1}^{N}\rho^{\rm GP}(\x_k)
\left(|\nabla_i F|^2+\suli_{j=1}^{i-1} v(|\x_i-\x_j|)|F|^2-8\pi g
\rho^{\rm GP}(\x_i)|F|^2\right)} {\int\prod_{k=1}^{N}\rho^{\rm
GP}(\x_k)|F|^2}.
\end{equation}
Compared to the expression for the energy involving $\Psi$ itself
we have thus obtained the replacements
\begin{equation}\label{repl}
V(\x)\to -8\pi g\rho^{\rm GP}(\x) \quad\mbox{and}\quad
\prod_{i=1}^Nd\x_i \to \prod_{i=1}^N\rho^{\rm GP}(\x_{i})d\x_{i}
\end{equation}
(recall that $\rho^{\rm GP}(\x)=|\Phi^{\rm GP}(\x)|^2$). We now
use the box method on {\it this} problem. More precisely, labeling
the boxes by an index $\alpha$, we have
\begin{equation}
\inf_F Q(F)\geq \inf_{\{n_\al\}} \suli_\al \inf_{F_\al}Q_\al (F_\al),
\end{equation}
where $Q_\al$ is defined by the same formula as $Q$  but with the
integrations limited to the box $\alpha$,  $F_{\alpha}$ is a wave
function with particle number $n_\alpha$, and the infimum is taken
over all distributions of the particles with $\sum n_\al=N$.
Approximating $\rho^{\rm GP}$ by a constant in each box, we can
use the bound on the homogeneous case (\cite{LY1998} and
\cite{LY2000}) in each box. To control the error terms, we need a
lower bound on the ratio $\rmin/\rmax$ in each box, which stems
from the measures in (\ref{ener3}), where $\rmax$ and $\rmin$,
respectively, denote the maximal and minimal values of $\rho^{\rm
GP}$ in box $\al$.

The problem is that for any fixed size of boxes ${\rmin}/{\rmax}$
tends rapidly to zero for boxes far from the origin. This problem
can be solved by enclosing the whole system in a big box
$\Lambda_{R}$ of side length $R$, with Neumann conditions on the
boundary. Replacing $\Phi^{\rm GP}$ by $\Phi^{\rm GP}_R$, which is
the minimizer of $\E^{\rm GP}$ restricted to $\Lambda_R$ (and
satisfies Neumann conditions), and restricting the integrations to
$\Lambda_{R}$ we can let the side lengths of the small boxes tend
to zero and be sure that ${\rmin}/{\rmax}\to 1$ uniformly for all
the boxes $\alpha$. However, we must control the error made by
enclosing the system in the big box. Let $E^{\rm QM}_{R}(N,a)$
denote the quantum mechanical ground state energy in the box
$\Lambda_R$. The essential step is
\begin{lemma}\label{lem1}
There is an $R_{0}<\infty$, depending only on $Ng$, such that
\begin{equation}\label{err}
    E^{\rm QM}(N,a)\geq E^{\rm QM}_{R}(N,a)
\end{equation}
for all $R\geq R_{0}$ and all $N$, $a$ with $Ng$ fixed.
\end{lemma}
This lemma follows from $V(\x)\to\infty$ for $|\x|\to\infty$,
together with an estimate for the chemical potential
\begin{equation}
E^{\rm QM}_{R}(N+1,a)-E^{\rm QM}_{R}(N,a)\leq e(Ng) (1+o(1))
\label{49}
\end{equation}
where $e(Ng)$ depends only on $Ng$ and is independent of $R$. The
proof of (\ref{49}) is similar to the proof of the upper bound
(\ref{ubd}).

Now $\rmin/\rmax$ is bounded below uniformly in each small box
contained in $\Lambda_R$. We first let the size of the boxes tend
to zero as $N\to\infty$, and finally take the limit $R\to\infty$.
This proves the desired lower bound (for fixed $Ng$, i.e. the GP
case).

If $Ng\to\infty$ as $N\to\infty$ the method above does not work
since the $R_0$ of Lemma \ref{lem1} tends to infinity with $Ng$.
However, using the explicit form of the TF minimizer, namely
\begin{equation}\label{tfminim}
\rho^{\rm TF}_{N,g}(\x)=\frac 1{8\pi g}[\mu^{\rm TF}-V(\x)]_+,
\end{equation}
where  $[t]_+\equiv\max\{t,0\}$ and $\mu^{\rm TF}$ is chosen so
that the normalization condition $\int \rho^{\rm TF}_{N,g}=N$
holds, we can use $V(\x)\geq \mu^{\rm TF}-8\pi g \rho^{\rm
TF}(\x)$ to get an replacement as in (\ref{repl}) without changing
the measure. Moreover, $\rho^{\rm TF}$ has compact support, so,
applying again the box method described above, the boxes far out
do not contribute to the energy. However, $\mu^{\rm TF}$ (which
depends only on the combination $Ng$) tends to infinity as
$Ng\to\infty$. Therefore we need to control the asymptotic
behaviour of the external potential, and this leads to the restrictions
on $V$ described in the paragraph preceding Theorem \ref{thm2}.

After controlling all error terms properly, we arrive at the
result
\begin{equation}\label{lowertf}
\liminf_{N\to\infty}\frac{E^{\rm QM}(N,a)}{E^{\rm TF}(N,g)}\geq 1
\end{equation}
in the limit $N\to\infty$, $a^D\bar\rho\to 0$ and $Ng\to\infty$.
Together with the upper bound (\ref{ubd}) and the fact that
$E^{\rm GP}(N,g)/E^{\rm TF}(N,g)=E^{\rm GP}(1,Ng)/E^{\rm
TF}(1,Ng)\to 1$ as $Ng\to\infty$ this proves Theorem \ref{thm2}.

%------------------------------------------------------------
% Appendix
%------------------------------------------------------------

%\appendix
%\section{Appendix}

%------------------------------------------------------------
% Acknowledgements (Optional)
%------------------------------------------------------------

%\begin{acknowledge}

%\end{acknowledge}

%------------------------------------------------------------
% References
% Use initial(s) of the author(s) for labeling.
%------------------------------------------------------------

\end{document}